\documentclass[conference]{IEEEtran}
\IEEEoverridecommandlockouts

\usepackage{times}
\usepackage{epsfig}
\usepackage{graphicx}
\usepackage{amsmath}
\usepackage{amssymb}
\usepackage[overload]{empheq}
\usepackage{subcaption}
\usepackage{multirow}
\usepackage{xcolor}

\usepackage[pagebackref=true,breaklinks=true,letterpaper=true,colorlinks,bookmarks=false]{hyperref}

\newcommand{\vect}[1]{\underline{\mathbf{#1}}}
\newcommand{\ie}{\textit{i.e.} }

\begin{document}

\title{Physically Plausible Spectral Reconstruction from RGB Images}

\author{Yi-Tun Lin\\
University of East Anglia\\
{\tt\small Yi-Tun.Lin@uea.ac.uk}
\and
Graham D. Finlayson\\
University of East Anglia\\
{\tt\small g.finlayson@uea.ac.uk}
}

\maketitle

\begin{abstract}

Recently Convolutional Neural Networks (CNN) have been used to reconstruct hyperspectral information from RGB images. Moreover, this \textbf{spectral reconstruction} problem (SR) can often be solved with good (low) error. However, these methods are not physically plausible: that is when the recovered spectra are reintegrated with the underlying camera sensitivities, the resulting predicted RGB is not the same as the actual RGB, and sometimes this discrepancy can be large. The problem is further compounded by exposure change. Indeed, most learning-based SR models train for a fixed exposure setting and we show that this can result in poor performance when exposure varies.

In this paper we show how CNN learning can be extended so that physical plausibility is enforced and the problem resulting from changing exposures is mitigated. Our SR solution improves the state-of-the-art spectral recovery performance under varying exposure conditions while simultaneously ensuring physical plausibility (\ie the recovered spectra reintegrate to the input RGBs exactly).

\end{abstract}

\section{Introduction}

Hyperspectral imaging devices are developed to capture high resolution radiance spectra at every pixel in an image, namely the \textit{hyperspectral images}.
These images often record additional scene information that are `invisible' to human eyes and consumer RGB cameras (where the spectral information is recorded with only 3 intensity values per pixel), which has been found useful in numerous computer vision applications including remote sensing \cite{veganzones2014hyperspectral,chen2015spectral,ghamisi2014survey,tao2015unsupervised,chen2014spectral}, anomaly detection \cite{jablonski2015principal} and medical imaging \cite{zhang2016tensor,zhang2016spectral}, as well as computer graphics applications such as scene relighting \cite{lam2013spectral} and digital art archiving \cite{xu2017self}.

Recent development in hyperspectral technology seeks faster image captures comparing to the conventional scanning-based techniques \cite{gat2000imaging,green1998imaging}. Several attempts have been made for real-time multi-channel capturing \cite{cao2011prism,wang2015high,oh2016yourself,takatani2017one}.
However, these devices are complicated and/or bulky that limits their usefulness. 
Other designs deploy novel optical components with specialized post-processing algorithms \cite{correa2015snapshot,garcia2018multi,arguello2014colored,galvis2017coded,lin2014spatial,rueda2015dmd,zhao2019hyperspectral}. But, these devices trade off spatial resolution and/or light sensitivity for faster capturing speed.

\begin{figure}
    \centering
    \includegraphics[width=0.9\linewidth]
                    {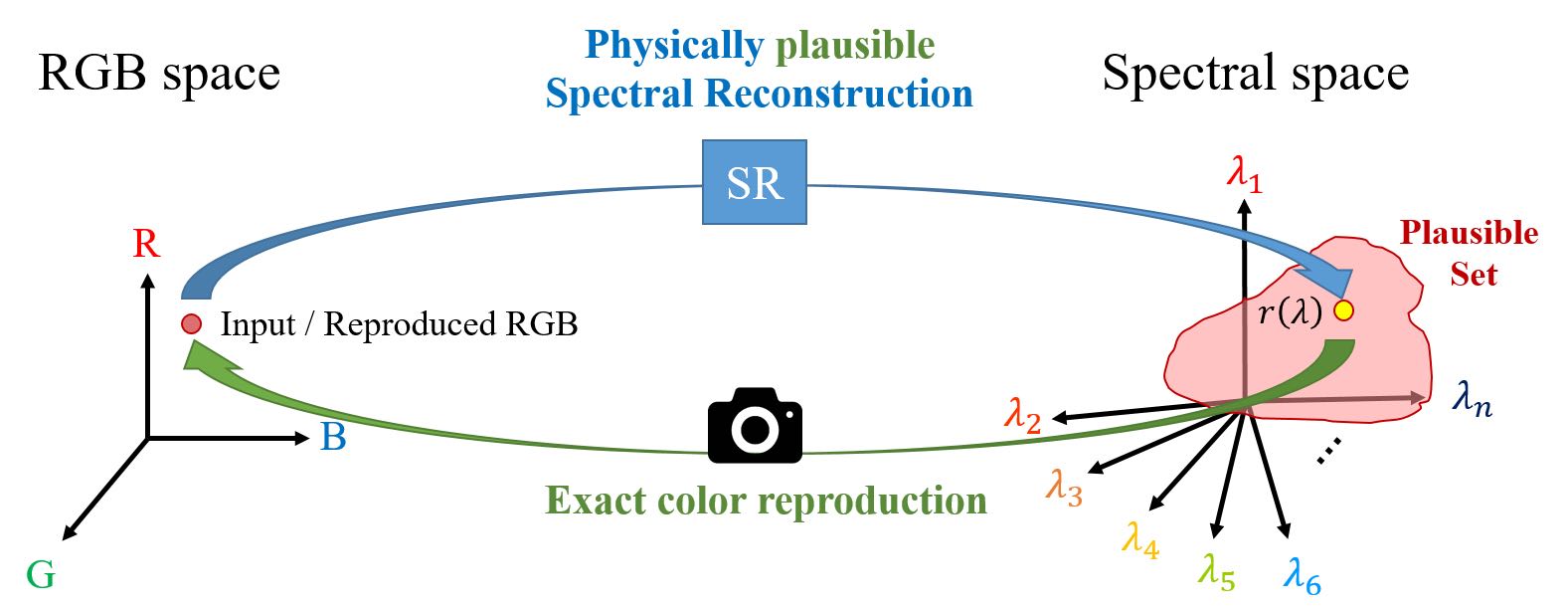}
    \includegraphics[width=0.9\linewidth]
                    {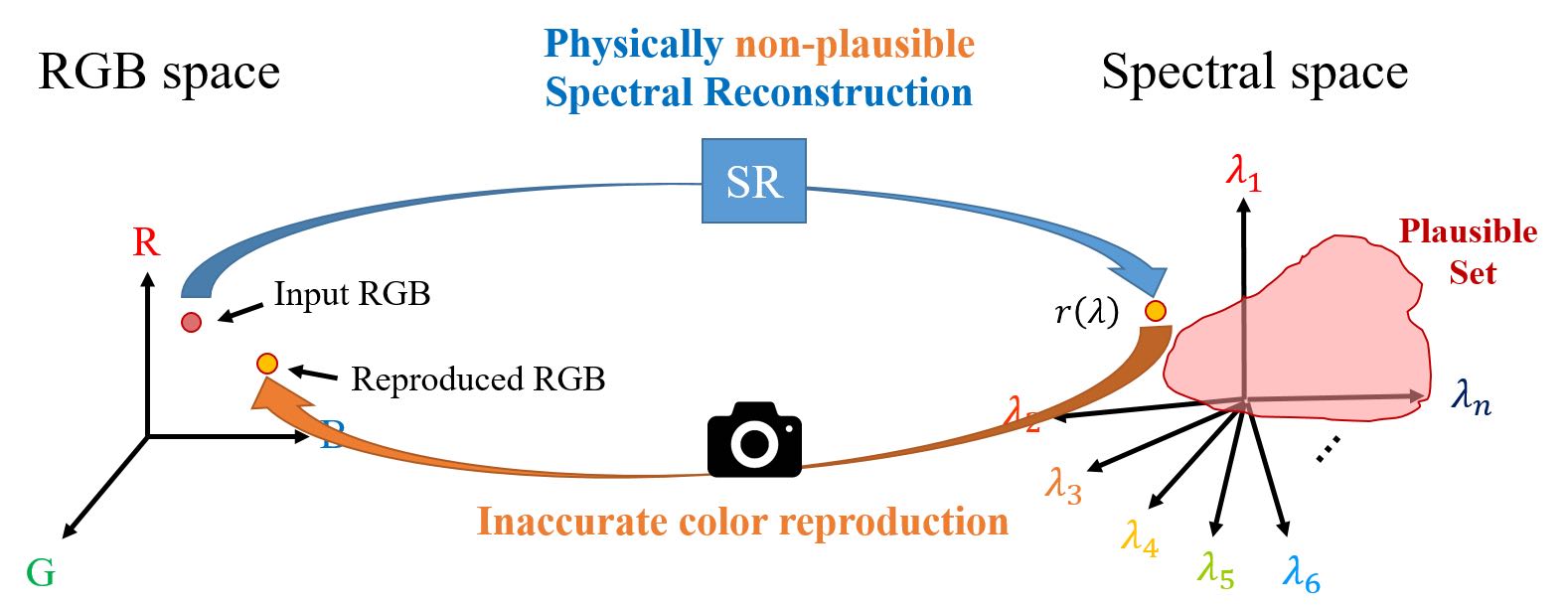}                
    \caption{Physically plausible spectral reconstruction (top) and non-plausible spectral reconstruction (bottom).}
    \label{fig:metamer_set}
\end{figure}

Spectral reconstruction (SR) is an alternative approach to recording hyperspectral information, where hyperspectral images are recovered from RGB images \cite{nguyen2014training,alvarez2017adversarial,liang2017optimized,koundinya20182d,stiebel2018reconstructing,fu2018joint,cao2017spectral,liang2017optimized,heikkinen2008evaluation,arad2016sparse,aeschbacher2017defense,fu2018joint,shi2018hscnn+,arad2018ntire}. The idea is not as na\"{i}ve as it might first appear. Indeed, we are expecting an RGB, which has just 3 numbers, to recover much more than 3 degrees of freedom in spectra. 
Fortunately, in natural scenes significant portion of the spectral variation is covered by its color appearance (\ie the RGBs) \cite{chakrabarti2011statistics}, which makes it possible for learning approaches to give rather accurate spectral approximations.
Recent approaches, leading by Convolutional Neural Networks (CNN), incorporate the images' spatial context to further enhance the accuracy of spectral recovery.

A key concern of this paper is the \textit{physical plausibility} of the spectral reconstruction algorithms. If we physically measure the radiance spectra by an accurate hyperspectral camera, given the 3 spectral sensitivity functions of an RGB camera, we are guaranteed to produce the RGBs which the RGB camera will actually give. 
Unfortunately most - and all deep neural network based - spectral reconstruction algorithms do not ensure this property. Indeed, as we shall show in this paper, the predicted RGB can be quite far from the actual one. This is not just `unfortunate' but completely missing out one of the key reasons we would like to use the spectral measurements: to \textit{predict} what we see, for example to better predict our own color sensation when the viewing conditions change (\ie the \textit{color correction} problem).

In Figure \ref{fig:metamer_set} we illustrate the idea of `physical plausibility'. The physically plausible spectral reconstruction is illustrated in the top diagram. Here a radiance spectra (\ie a spectral power distribution   $r(\lambda)$, see the right side of the image) is recovered from an RGB (the red point on the left in RGB space). Now we reintegrate the spectra with the camera's spectral sensitivities - simulate taking a picture of this spectrum - which gives a predicted RGB. In this case the input RGB and the predicted counterpart are the same.

The diagram in the bottom half of Figure \ref{fig:metamer_set} shows a spectral recovery which produces incorrect color when reintegrated with the camera sensitivities. This `physically non-plausible spectral reconstruction' is the norm (and is a feature exhibited by all deep network based algorithms we are aware of). Put bluntly, these algorithms provide the estimations of spectra which - because they do not reintegrate to the input RGB - {\bf must be the wrong answers}.

\begin{figure}
    \centering
    \includegraphics[width=0.9\linewidth]{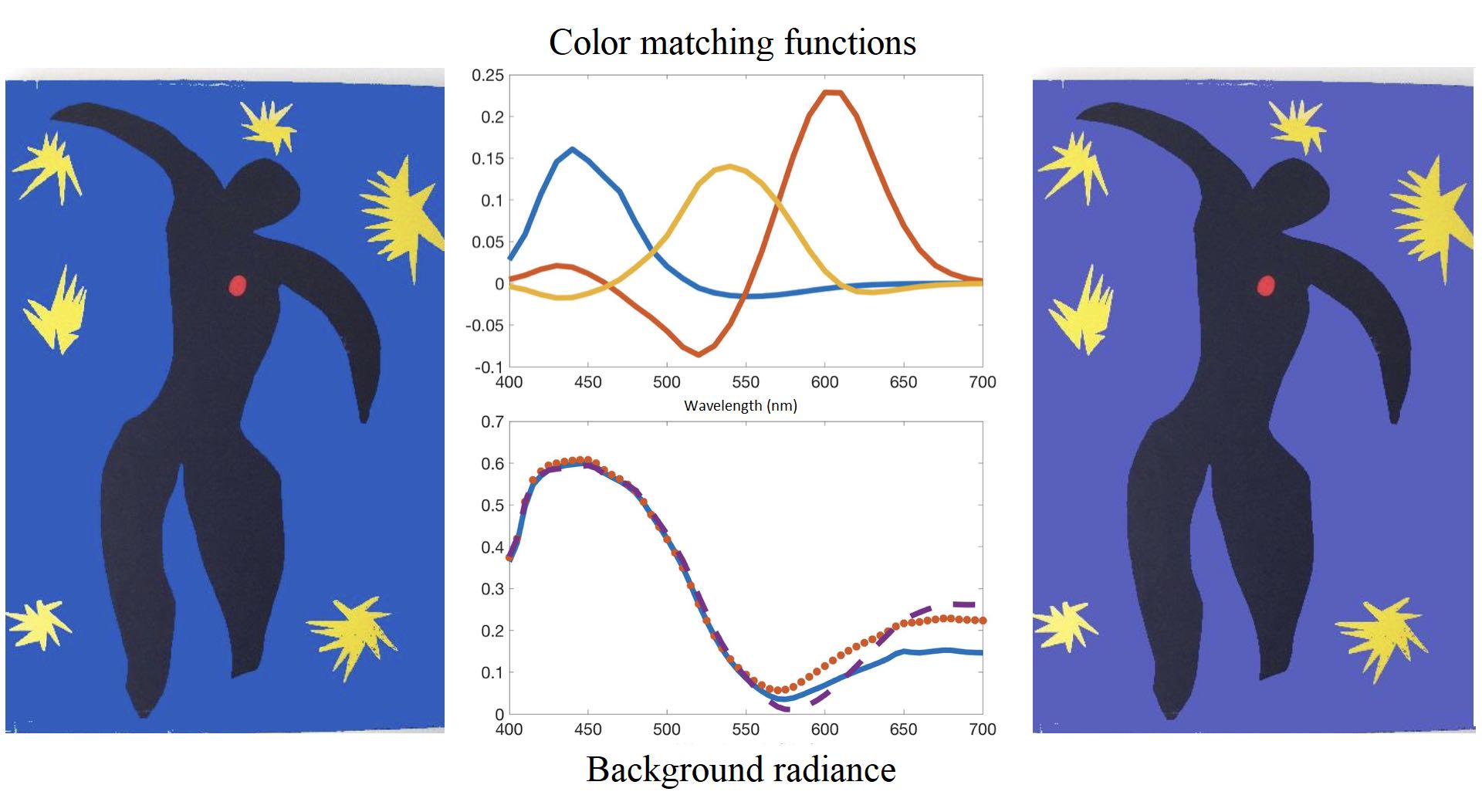}
    \caption{The color fidelity test on the artist Matisse's famous painting `Jazz' for two reconstructed spectra with similar spectral error.}
    \label{fig:jazz}
\end{figure}

In Figure \ref{fig:jazz} we show a pictorial example. In the bottom-middle panel we show two reconstructed spectra - red and purple dotted curves - having similar spectral difference from the ground-truth (blue solid curve). By integrating the sRGB display color matching functions given in the middle-top panel, the purple curve reproduces the background color exactly as the original painting on the left. In contrary, the red curve reproduces the image on the right, which shows significant background color shift. 

The examples illustrated in Figures \ref{fig:metamer_set} and \ref{fig:jazz} show the advantage of incorporating the physics of image formation into learning-based methods for spectral reconstruction. 
Another related issue we also consider in this paper is \textit{exposure invariance}. Clearly, if the light intensity in the scene changes, the ground-truth radiance spectra will be linearly scaled, and for linear RGB images (\ie the camera raw data), the RGBs will also be scaled in the same way. However, as we later show in this paper this second physical reality is also not preserved in the state-of-the-art CNN models: these networks were trained for a single exposure condition and they perform poorly when a different exposure setting is tested.

This paper makes three main contributions:
\begin{itemize}
\itemsep0em
    \item We evaluate the state-of-the-art HSCNN-D and HSCNN-R models \cite{arad2018ntire,shi2018hscnn+} to gauge the extent that they deliver physically plausible spectral recovery, either in the sense of predicting the input RGBs or being resilient to a varying exposure. 
    \item We propose a novel framework which ensures \textit{exact} color reproduction in CNN-based spectral reconstruction.
    \item We design a data augmentation process that maintains model stability over different exposure settings.
\end{itemize}

The rest of the paper is organized as follows. In section 2 we review the related field. In section 3 we show how we can solve the spectral reconstruction problem while ensuring physical plausibility. Implementation details are given in section 4. Experimental results are presented in section 5. The paper concludes in section 6.

\section{Related Work}

\textbf{Hyperspectral imaging.} 
There exist technologies where hyperspectral images can be directly captured, and these include using a prism-mask system \cite{cao2011prism}, multiple cameras \cite{wang2015dual,oh2016yourself} and faced reflectors \cite{takatani2017one}. However, the practical applciation of these devices is limited by their complex configurations and/or their physical bulkiness.
Alternately, in  \textit{compressive imaging}, a scene's spectral information is encoded in alternative forms on the sensed 2-D images. But, there is the overhead of decompressing the signal.
Examples include multi-spectral color filter array \cite{correa2015snapshot}, coded aperture \cite{arguello2014colored,galvis2017coded,garcia2018multi}, diffractive gratings \cite{lin2014spatial}, digital micro-mirror device \cite{rueda2015dmd} and most recently random printed mask \cite{zhao2019hyperspectral}.
Other problems inherent in compressive sensing are the need for specialized optics and the inherent trade-off between the number of sensors and/or light sensitivity and the spatial resolution.

\textbf{Spectral reconstruction (SR).} 
Rather than building new hardware for capturing hyperspectral images, spectral reconstruction attempts to map RGB images to their spectral counterparts. Shallow-learned methods - of which sparse coding is the best example \cite{arad2016sparse,aeschbacher2017defense} - have the advantage of model simplicity and quick training. However, these models is effectively implementing a `one-to-one' lookup table, which contradicts the fact that \textit{many} (in fact \textit{infinite}) spectra can reproduce the same RGB. 

In the CNN approach the implementation complexity is much higher as so the hardware requirements but the reconstruction is richer. The promise of these methods is that, in an intermediate representation, they might identify scene contents which are associated with the target spectra and then effectively use these information in the recovery process. Indeed, it is well known that faces, chlorophyl (in foliage) and daylights have very characteristic shapes (amongst other scene features). Of the current developments, deep neural networks \cite{alvarez2017adversarial,koundinya20182d,stiebel2018reconstructing,fu2018joint,shi2018hscnn+} provide the leading performance in spectral reconstruction.

\textbf{Physical plausibility.} In this paper we address the importance that, as an alternative way of getting hyperspectral information, a spectral reconstruction algorithm should always produce physically plausible radiance predictions which can be reintegrated to the same RGB values as they are recovered from. 

Interestingly, some of the early models can already provide accurate color reproduction (but much poorer spectral recovery comparing to the recent CNN methods). 
For example, using weighted-PCA based on color differences \cite{agahian2008reconstruction} and
colorimetrically correcting the linear regression spectral recovery (\ie pseudo-inverse) \cite{zhao2007image}. Furthermore, sparse coding methods \cite{arad2016sparse,aeschbacher2017defense} can also provide rather accurate color reproduction by virtue of their fundamental `neighbor embedding' assumption \cite{timofte2014a+}. Finally, the complex and computationally laborious Bayesian inference method \cite{morovic2006metamer} was also introduced where physical plausibility is ensured. 

However, in the recent NTIRE 2018 Challenge on Spectral Reconstruction from RGB Images (hereinafter abbreviated as NTIRE2018) \cite{arad2018ntire}, all 12 leading entries out of 73 attendants (on the `Clean Track') involve the implementation of deep neural networks. None of these methods explicitly ensure the spectra can reintegrate to the input RGBs. 

\textbf{Exposure invariance.}
In many learning-based computer vision tasks, the model stability over intensity change are considered; that is, the model are ensured to work well even as the scene exposure changes. 
However, Lin and Finlayson \cite{lin2019exposure} demonstrated that leading spectral reconstruction models in NTIRE2018 perform poorly in different exposure settings, and this has raised a concern that many modern developments of spectral reconstruction may not work in the wild where exposure can vary. 

\section{Physically Plausible Spectral Reconstruction}

At each pixel of a hyperspectral image, a high-resolution radiance spectrum is recorded. The corresponding RGB image is simulated by calculating the \textit{inner products} between the measured radiance spectra and the spectral sensitivity functions of the RGB camera:
\begin{subequations}
\label{eqn:image_formation}
\begin{equation}
    \rho_k = \sum_{\lambda\in\Omega} s_k(\lambda)r(\lambda)\ ,
\end{equation}
where $k = 1,2,3$ refer to the red, green and blue channels of the RGB image, $\rho_k$, $s_k(\lambda)$ and $r(\lambda)$ are respectively the $k$-th camera response, the $k$-th camera sensitivity function and the radiance function, $\lambda$ denotes the wavelength dimension, and $\Omega$ is the \textit{visible spectrum}. Of course for this inner-product model (as oppose to an integral) of image formation to work, we must sample the spectra at a sufficient resolution across the visible spectrum. In all simulations we report later in this paper, we assume the visible spectrum runs from 400 through 700 nanometers, and the spectra are sampled every 10 nanometers (this is the common assumption made in most studies, including the NTIRE2018 \cite{arad2018ntire}).

Let us  vectorize the above equation:
\begin{equation}
    \vect{\rho} = \mathbf{S}^\mathsf{T}\vect{r}\ ,
\end{equation}
\end{subequations}
where $\vect{\rho} = (\rho_1,\rho_2,\rho_3)^\mathsf{T}$ is the $3$-dimensional RGB vector, $\vect{r}$ is the $n$-dimensional radiance spectra with $n$ to be the number of spectral bands, and $\mathbf{S} = (\vect{s}_1, \vect{s}_2, \vect{s}_3)$ is an $n\times 3$ matrix with its columns to be the three distinct camera sensitivity functions.

In the ordinary spectral reconstruction framework, the radiance spectrum $\vect{r}$ is recovered from the RGB camera response $\vect{\rho}$: the spectral reconstruction algorithm searches for the best solution to $\vect{r}$ within the entire spectral space (\ie $\mathbb{R}^n$) that statistically minimizes the distance error between the recovered and ground-truth radiance spectra. However, this framework does not ensure that the reconstructed $\vect{r}$ must reproduce $\vect{\rho}$ - the algorithm may find a solution which is spectrally close to the ground-truth but reproduces distant color (as per the example we showed in Figure \ref{fig:jazz}).

Let us now develop a method to \textit{constrain} the algorithm only to search for the estimated radiance within the set of spectra that integrates to the correct RGB. For this purpose, we propose a \textit{plausible set} concept, which is defined as the set of all spectra that integrate to a target RGB.

The derivation of our plausible set is analogous to, but simpler than, the \textit{metamer set} in \cite{finlayson2005metamer,morovic2006metamer}, while their focus was on the reflectance set instead of our case on the radiance set.

\subsection{The Plausible Set}

Given known camera sensitivity functions $\mathbf{S}$, the plausible set $\mathcal{P}$ is defined as:
\begin{equation}
    \mathcal{P}(\vect{\rho};\mathbf{S}) = \bigg\{ \vect{r}\ \bigg|\ \mathbf{S}^\mathsf{T}\vect{r} = \vect{\rho} \bigg\}\ .
\end{equation}
Geometrically, the outcome of an inner product is only affected by the parts of the two vectors that are `parallel' to each other, whereas the `perpendicular' part do not contribute to the product.

Given this view, the constraint $\mathbf{S}^\mathsf{T}\vect{r}=\vect{\rho}$ in effect separates $\vect{r}$ into two parts: the part that is spanned by the column vectors of $\mathbf{S}$ which contributes to $\vect{\rho}$, and the part lies in the \textit{null-space} of $\mathbf{S}$ which yields \textit{zero projection}. That is,
\begin{equation}
\label{eqn:funda_metablack_separation}
    \vect{r} = \vect{r}^{\parallel} + \vect{r}^{\bot}\ ,
\end{equation}
subject to
\begin{subequations}
\label{eqn:funda_metablack}
\begin{align}[left = \empheqlbrace\, ]
    \mathbf{S}^\mathsf{T}\vect{r}^{\parallel} &= \vect{\rho}\\
    \mathbf{S}^\mathsf{T}\vect{r}^{\bot} &= \vect{0}\ \ \ 
\end{align}
and
\begin{equation}
    \vect{r}^\parallel \cdot\  \vect{r}^\bot = 0\ .
\end{equation}
\end{subequations}

The $\vect{r}^\parallel$ component can be derived directly by the subspace projection. The projection matrix with respect to $\mathbf{S}$ is written as:
\begin{equation}
\label{eqn:projection_S}
    \mathbf{P}^{\mathbf{S}} = \mathbf{S}(\mathbf{S}^\mathsf{T}\mathbf{S})^{-1}\mathbf{S}^\mathsf{T}\ , 
\end{equation}
such that
\begin{equation}
    \vect{r}^{\parallel} = \mathbf{P}^\mathbf{S}\vect{r} = \mathbf{S}(\mathbf{S}^\mathsf{T}\mathbf{S})^{-1}\mathbf{S}^\mathsf{T}\vect{r}\ .
\end{equation}
Next, we enforce our desired constraint $\mathbf{S}^\mathsf{T}\vect{r} = \vect{\rho}$ on $\vect{r}^\parallel$, which gives
\begin{equation}
\label{eqn:fundamental_solution}
    \vect{r}^\parallel = \mathbf{S}(\mathbf{S}^\mathsf{T}\mathbf{S})^{-1}\vect{\rho}\ .
\end{equation}
It is important to see that the derived $\vect{r}^\parallel$ is \textit{fixed} given $\mathbf{S}$ and $\vect{\rho}$, which implies that all $\vect{r} \in \mathcal{P}(\vect{\rho};\mathbf{S})$ shares the same $\vect{r}^\parallel$ and only the $\vect{r}^\bot$ component determines the difference between the radiance in set $\mathcal{P}$.

\begin{figure*}[h]
    \centering
    \includegraphics[width=1\linewidth]{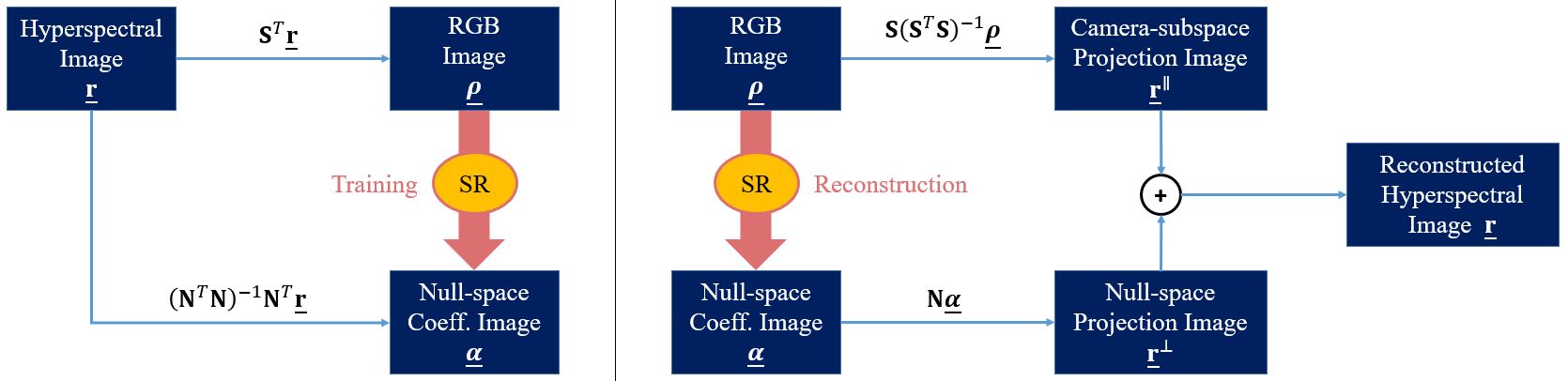}
    \caption{The training (left) and reconstruction scheme (right) of our physically plausible spectral reconstruction.}
    \label{fig:traing_testing}
\end{figure*}

On the other hand, $\vect{r}^\bot$ can be \textit{any} vector in the null space of $\mathbf{S}$, which is spanned by $n-3$ linearly independent bases that are orthogonal to all $3$ column vectors of $\mathbf{S}$. This set of bases can be obtained by finding the non-trivial solutions of $\vect{r}^\bot$ in Equation (\ref{eqn:funda_metablack}b), or by calculating the  $n-3$ basis vectors  that span ${\cal I}_{n\times n}-\mathbf{P}^{\mathbf{S}}$ (${\cal I}_{n\times n}$ is the $n\times n$ identity matrix), which is the projection matrix with respect to all other dimensions in $\mathbb{R}^n$ that are orthogonal to the column space of $\mathbf{S}$.
In either way, we get a set of $n-3$ null-space bases, and every $\vect{r}^\bot$ can be uniquely derived by a linear combination of them:
\begin{equation}
\label{eqn:metablack_by_null-space_coeff}
    \vect{r}^\bot = \mathbf{N}\vect{\alpha}\ ,\quad \vect{\alpha} \in \mathbb{R}^{n-3}\ ,
\end{equation}
where $\mathbf{N}$ is an $n\times n-3$ matrix with its columns to be the null-space basis vectors. We call $\vect{\alpha}$ the \textit{null-space coefficients}. Finally, we reach the following definition of $\mathcal{P}$:
\begin{equation}
\label{eqn:plausible_set_final}
    \mathcal{P}(\vect{\rho};\mathbf{S}) = \bigg\{ \mathbf{S}(\mathbf{S}^\mathsf{T}\mathbf{S})^{-1}\vect{\rho} + \mathbf{N}\vect{\alpha}\ \bigg|\ \vect{\alpha} \in \mathbb{R}^{n-3} \bigg\}\ ,
\end{equation}
where all $\mathbf{S}$, $\mathbf{N}$ and $\vect{\rho}$ are known factors, leaving $\vect{\alpha}$ to be the only variation within the plausible set $\mathcal{P}$.

In the next part of this section, we are going to introduce our physically plausible spectral reconstruction via the null-space coefficients reconstruction. 

\subsection{Reconstructing the Null-space Coefficients}

Recall Equation (\ref{eqn:image_formation}a) and (\ref{eqn:image_formation}b), given a ground-truth radiance $\vect{r}_\text{gt}$, the corresponding RGB is calculated by $\vect{\rho}_\text{gt} = \mathbf{S}^\mathsf{T}\vect{r}_\text{gt}$. This indicates that $\vect{r}_\text{gt}$ is a member of $\mathcal{P}(\vect{\rho}_\text{gt};\mathbf{S})$, in which it corresponds to one unique $\vect{\alpha}$, denoted as $\vect{\alpha}_\text{gt}$. Based on which, we further translate our goal of making the spectral reconstruction algorithm search for the reconstruction $\vect{r}_\text{rec}$ in $\mathcal{P}(\vect{\rho}_\text{gt};\mathbf{S})$, into seeking the null-space coefficients $\vect{\alpha}_\text{rec}$ in $\mathbb{R}^{n-3}$ which best approximates $\vect{\alpha}_\text{gt}$. Training the spectral reconstruction algorithm $\mathcal{SR}:\ \mathbb{R}^3 \mapsto \mathbb{R}^{n-3}$ such that
\begin{equation}
    \vect{\alpha}_\text{rec} = \mathcal{SR}(\vect{\rho}_\text{gt}) \approx \vect{\alpha}_\text{gt}\ ,
\end{equation}
the reconstructed spectrum $\vect{r}_\text{rec} \in \mathcal{P}(\vect{\rho}_\text{gt};\mathbf{S})$ is then derived by
\begin{equation}
    \vect{r}_\text{rec} = \mathbf{S}(\mathbf{S}^\mathsf{T}\mathbf{S})^{-1}\vect{\rho}_\text{gt} + \mathbf{N} \bigg( \mathcal{SR}(\vect{\rho}_\text{gt}) \bigg)\ .
\end{equation}

So far, the idea behind our physically plausible framework for spectral reconstruction has been established. Still, for CNN models the ground-truth labels are necessary, which means we are yet to calculate $\vect{\alpha}_\text{gt}$ from the $\vect{r}_\text{gt}$ in the hyperspectral images.

In Equation (\ref{eqn:projection_S}) we calculated the projection matrix $\mathbf{P}^\mathbf{S}$ which projects $\vect{r}$ onto the column space of $\mathbf{S}$ that derives $\vect{r}^\parallel$. Likewise, to derive $\vect{r}^\bot$ we seek the projection of $\vect{r}$ onto the column space of the $\mathbf{N}$. The null-space projection matrix can be written as:
\begin{equation}
    \mathbf{P}^{\mathbf{N}} = \mathbf{N}(\mathbf{N}^\mathsf{T}\mathbf{N})^{-1}\mathbf{N}^\mathsf{T}
\end{equation}
(which is equivalent to ${\cal I}_{n\times n}-\mathbf{P}^{\mathbf{S}}$), such that
\begin{equation}
\label{eqn:null_projection_by_basis}
    \vect{r}^\bot = \mathbf{P}^\mathbf{N}\vect{r} = \mathbf{N}(\mathbf{N}^\mathsf{T}\mathbf{N})^{-1}\mathbf{N}^\mathsf{T}\vect{r}\ .
\end{equation}
Together with Equation (\ref{eqn:metablack_by_null-space_coeff}), we get 
\begin{equation}
\label{eqn:null_projection_and_alpha}
    \mathbf{N}\vect{\alpha} = \mathbf{N} \bigg( (\mathbf{N}^\mathsf{T}\mathbf{N})^{-1}\mathbf{N}^\mathsf{T}\vect{r} \bigg)\ .
\end{equation}
Finally, since the columns of $\mathbf{N}$ (\ie the null-space basis vectors) are \textit{linearly-independent}, we derive
\begin{equation}
\label{eqn:alpha_equals_something_r}
    \vect{\alpha} = (\mathbf{N}^\mathsf{T}\mathbf{N})^{-1}\mathbf{N}^\mathsf{T}\vect{r}\ .
\end{equation}

Figure \ref{fig:traing_testing} summarizes the framework of our physically plausible spectral reconstruction. In the reconstruction stage, the camera-subspace projection $\vect{r}^\parallel$ is calculated directly from the RGB input and the spectral reconstruction algorithm only concerns the recovery of the null-space projection $\vect{r}^\bot$. As the color reproduction of the reconstructed hyperspectral image only depends on $\vect{r}^\parallel$ (Equation (\ref{eqn:funda_metablack}a) and (\ref{eqn:funda_metablack}b)), the reconstructed hyperspectral image is ensured to reproduce exactly the input RGB image. In the next section we are going to integrate this framework with the state-of-the-art spectral reconstruction model based on CNN.

\section{Implementation}

\begin{figure*}[h]
    \centering
    \includegraphics[width=1\linewidth]{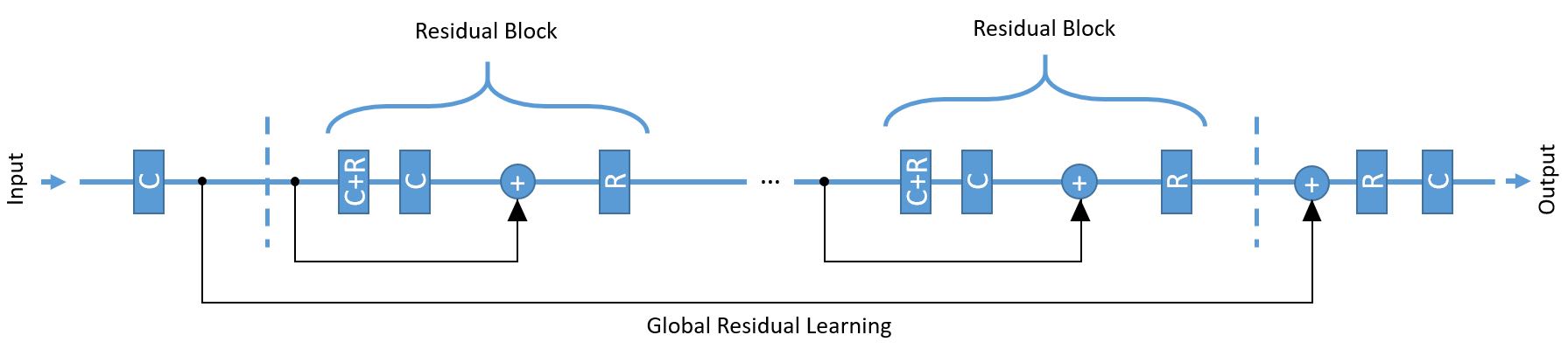}
    \caption{The HSCNN-R architecture \cite{shi2018hscnn+}. `C' means $3\times 3$ convolution and `R' refers to ReLU activation.}
    \label{fig:HSCNN-R}
\end{figure*}

We build our models based on the HSCNN-R architecture, which is the 2nd place entry of the NTIRE2018 \cite{arad2018ntire,shi2018hscnn+} (whose performance is similar to the 1st place HSCNN-D model; we use the 2nd place architecture simply because it was simpler in our development environment).
As illustrated in Figure \ref{fig:HSCNN-R}, the HSCNN-R model adopts a deep residual learning framework \cite{he2016deep}. Each of the residual blocks is constructed with two convolutional layers and one ReLU layer. The model also adopts a global residual learning structure. All convolutional kernels are set to $3\times 3$. 

On training and reconstruction, the network maps $50 \times 50$ RGB image patches to the corresponding $31$-channel hyperspectral patches (the spectral dimension runs from 400 to 700 $nm$ with 10 $nm$ sampling intervals). The final image is decided by the reconstruction outcome of 3 HSCNN-R networks with different filter numbers in each layer (64, 256 and 256) and depths (34, 20, and 30). 

In this paper, we aim for two improvements on HSCNN-R: (1) perfect color reproduction and (2) robustness against exposure change. For the former, we integrate our physically plausible framework to HSCNN-R, and for the latter, we propose a new \textit{data augmentation} process. To study the effects of both improvements, 3 new models listed in Table \ref{tab:list_of_new_models} are trained.

\begin{table}[h]
    \centering
    \begin{tabular}{p{5em}|c|c}
        \multirow{2}{*}{$\quad\textbf{Model}$}   & Physically      & Data          \\
                              & Plausible       & Augmentation  \\
        \hline
        $\text{HSCNN-R}^p$    & V               &               \\
        $\text{HSCNN-R}^d$    &                 & V             \\
        $\text{HSCNN-R}^{pd}$ & V               & V                     
    \end{tabular}
    \caption{List of our new models}
    \label{tab:list_of_new_models}
\end{table}

\subsection{Physically Plausible HSCNN-R}

In the original HSCNN-R model, the output layer corresponds to a $50\times 50$ hyperspectral image patch with $31$ spectral dimensions. To accommodate our physically plausible framework in HSCNN-R$^p$ and HSCNN-R$^{pd}$, we reduce the spectral dimension from $31$ to $28$ in the output layer for recovering the image of null-space coefficients ($\vect{\alpha}$ is $(n-3)$-dimensional with $n=31$).

Unlike the original hyperspectral data which only contains positive values, the null-space coefficients $\vect{\alpha}$ allow \textit{negative} entries, and this is not permitted for the ReLU output layer in the HSCNN-R architecture. As a result, it is necessary to re-center the ground-truth $\vect{\alpha}$ such that the negative values are prevented. 

In our implementation, 
we found that empirically the entries of the ground-truth $\vect{\alpha}$ range between $-1$ and $1$. Hence, in the training stage of HSCNN-R$^p$ we adopt a re-centering:
\begin{equation}
    \vect{\tilde{\alpha}} = (\vect{\alpha} + 1)/2\ .
\end{equation}
The reverse function is used in the reconstruction stage to center the targeted $\vect{\alpha}$ back from $\vect{\tilde{\alpha}}$. As we will mention later the HSCNN-R$^{pd}$ model requires a different re-centering function due to the implementation of our data augmentation process.

The rest of the hyperparameters of HSCNN-R$^p$ are kept the same as the original HSCNN-R model \cite{shi2018hscnn+}. Our HSCNN-R$^p$ is expected to provide absolute color reproduction. However, as shown in \cite{lin2019exposure}, the original HSCNN-R is not robust against intensity change, and this implies that HSCNN-R$^p$ also may not perform well in spectral recovery when the testing exposure condition varies. 

\subsection{Intensity-scaling Data Augmentation}

We create the augmented data by simulating `brighter' and `dimmer' RGB images from the ground-truth hyperspectral images. Instead of generating all the new data before training the model, we draw different scaling constants in real time during training: all input image patches (and the same patch in different training epochs) are scaled differently, which allows the network to see more intensity variation in the data.

Furthermore, since a spectral reconstruction algorithm can be potentially implemented on an RGB camera, we want to especially ensure that, when adjusting the standard exposure settings in the RGB cameras (\ie the aperture size and shutter speed) the trained model performs equally well. We remind that these settings by convention follow \textit{geometric progressions}; more precisely, the available aperture sizes normally follows a sequential scaling change by $\sqrt{2}$, and the shutter speed is adjusted by a factor of 2 between adjacent modes. Based on this fact, we propose to draw the scaling constants $\xi$ from a uniform distribution on a \textit{log scale}:
\begin{equation}
    \log_\beta \xi \sim Uniform(-1,1)\ .
\end{equation}
In our implementation, we set $\beta = 10$ such that the scaling factor $\xi$ is bounded by $[\frac{1}{10},10]$. 

For comparison, we train another intermediate model, HSCNN-R$^d$, which only adopts the intensity-scaling data augmentation. This model is refined from the pre-trained HSCNN-R provided in \cite{shi2018hscnn+}, and all hyperparameters are kept the same.

\begin{table*}[t!]
    \centering
    \centerline{
    \begin{tabular}{c||c|c||c|c||c|c||c|c||c|c||c|c}
    \hline
    \multirow{3}{*}{\textbf{Model}} & \multicolumn{4}{c||}{Original exposure ($\xi = 1$)} & \multicolumn{4}{c||}{Half exposure ($\xi = 0.5$)} & \multicolumn{4}{c}{Double exposure ($\xi = 2$)} \\
    \cline{2-13}
    & \multicolumn{2}{c||}{$\Delta E$} & \multicolumn{2}{c||}{MRAE ($\times \text{10}^\text{-2}$)} & \multicolumn{2}{c||}{$\Delta E$} & \multicolumn{2}{c||}{MRAE ($\times \text{10}^\text{-2}$)} & \multicolumn{2}{c||}{$\Delta E$} & \multicolumn{2}{c}{MRAE ($\times \text{10}^\text{-2}$)} \\
    \cline{2-13}
                    & Mean & WC & Mean & WC & Mean & WC & Mean & WC & Mean & WC & Mean & WC \\
    \hline
    HSCNN-D         & 0.51 & 10.18 & \textcolor{red}{1.19} & \textcolor{red}{14.09} & 1.75 & 11.08 & 14.87 & 48.24 & 0.51 & 9.17 & 5.79 & \textcolor{blue}{23.36} \\
    HSCNN-R         & 0.49 & 13.10 & \textcolor{blue}{1.35} & 23.84 & 1.98 & 13.81 & 17.41 & 71.94 & 0.67 & 9.45 & 5.86 & 24.95   \\
    \hline\hline
    HSCNN-R$^{p\ }$ & \textcolor{red}{0.00} & \textcolor{red}{0.00} & 1.73 & \textcolor{blue}{19.31} & \textcolor{red}{0.00} & \textcolor{red}{0.00} & 14.23 & 37.94 & \textcolor{red}{0.00} & \textcolor{red}{0.00} & 6.43 & 24.84  \\
    HSCNN-R$^{d\ }$ & 0.26 & 8.98 & 2.77 & 19.78 & 0.25 & 8.92 & \textcolor{red}{2.78} & \textcolor{red}{19.83} & 0.26 & 9.07 & \textcolor{red}{2.77} & \textcolor{red}{19.73}   \\
    HSCNN-R$^{pd}$  & \textcolor{red}{0.00} & \textcolor{red}{0.00} & 2.80 & 23.93 & \textcolor{red}{0.00} & \textcolor{red}{0.00} & \textcolor{blue}{2.91} & \textcolor{blue}{23.89} & \textcolor{red}{0.00} & \textcolor{red}{0.00} & \textcolor{blue}{2.78} & 24.09  \\
    \hline
    \end{tabular}
    }
    \caption{The mean and the worst-case (WC) hyperspectral image reconstruction error in $\Delta E$ and MRAE under original, half and double exposure settings. Best results are shown in \textcolor{red}{red} and the second-best results are shown in \textcolor{blue}{blue}.}
    \label{tab:result}
\end{table*}

On the other hand, to apply this new data augmentation framework on the physically plausible model, we need to adjust the re-centering function to accommodate the change in range of the entries of $\vect{\alpha}$. In our case, as $\beta = 10$:
\begin{equation}
    \vect{\tilde{\alpha}} = (\vect{\alpha}+10)/20\ .
\end{equation}
Additionally, to make the model converge efficiently, we set the adaptive learning rate to follow a polynomial decay with the power of $25$ (instead of the original $1.5$). This final model is referred to as HSCNN-R$^{pd}$.

\section{Experiment}
\subsection{Experimental Setup}
We trained our new models (as listed in Table \ref{tab:list_of_new_models}) based on the ICVL database \cite{arad2016sparse} (201 hyperspectral images), where we randomly split the database into 100 images for training, 50 for validation and 50 for evaluation. The CIE 1964 color matching functions \cite{commission1964cie} were selected as the camera sensitivity functions, by which the ground-truth RGBs (\ie the CIEXYZ color coordinates) were simulated. We also tested the original HSCNN-D and HSCNN-R models (the pre-trained networks in \cite{shi2018hscnn+} were directly used) to compare with our new models.

Our experiment concerns the performances of the models in terms of (1) color reproduction, (2) spectral recovery and (3) both performances under different exposure settings. We select the following error metrics:
\begin{itemize}
    \item Color difference: CIE 1976 color difference\\
    \begin{equation}
    \label{eqn:deltaE}
        \Delta E = \sqrt{(L^*_\text{gt} - L^*_\text{rec})^2 + (a^*_\text{gt} - a^*_\text{rec})^2 + (b^*_\text{gt} - b^*_\text{rec})^2}
    \end{equation}
    \\
    \item Spectral difference: Mean Relative Absolute Error\\
    \begin{equation}
    \label{eqn:mrae}
        \text{MRAE} = \frac{1}{n} \bigg|\bigg| \frac{\vect{r}_\text{gt} - \vect{r}_\text{rec}}{\vect{r}_\text{gt}} \bigg|\bigg|_1
    \end{equation}
\end{itemize}

Equation (\ref{eqn:deltaE}) shows the definition of the CIE 1976 color difference formula \cite{robertson1977cie}, where $(L^*_\text{gt}, a^*_\text{gt}, b^*_\text{gt})$ and $(L^*_\text{rec}, a^*_\text{rec}, b^*_\text{rec})$ are the CIELAB color coordinates of the ground-truth and reconstructed RGB colors, respectively. The transformation between CIEXYZ and CIELAB requires the normalization by the `white point' coordinates (\ie the illumination color), for which we hand-craft the white points of each images by selecting the RGB of the brightest achromatic pixel. 

In Equation (\ref{eqn:mrae}), respectively $\vect{r}_\text{gt}$ and $\vect{r}_\text{rec}$ refers to the ground-truth and reconstructed radiance spectra, and $n$ is the number of spectral bands. The division is component-wise and the $L_1$ norm is calculated. 

Note that both of the above metrics are pixel-wisely defined, which means the performance of each pixel in an image is considered independently. In addition, since both metrics involve normalization of the reference intensity: for $\Delta E$ the illumination white-point coordinates are divided, and for MRAE the spectral difference is divided by the ground-truth spectrum. This ensures that our performance measurements are independent to the overall intensity of the compared targets.

We test all models under 3 exposure settings: the original, half and double exposure. For each testing exposure, we uniformly scale-up all the evaluation images with the same scaling constant (respectively $\xi$ = 1, 0.5 and 2), and the reconstructed hyperspectral images are compared with the ground-truth hyperspectral images scaled by the same constant.

\subsection{Result and Discussion}

The performance statistics are shown in Table \ref{tab:result}. We show the mean and the worst-case (WC) performance of the models. The `worst case' is defined per image as the averaged error of `the worst 1000 pixels' (the image dimension is around $1300 \times 1392$), and the worst-case performance given in Table \ref{tab:result} refers to the mean worst-case error over all evaluation set.

\begin{figure*}[h!]
    \centering
    \includegraphics[width=1\linewidth]{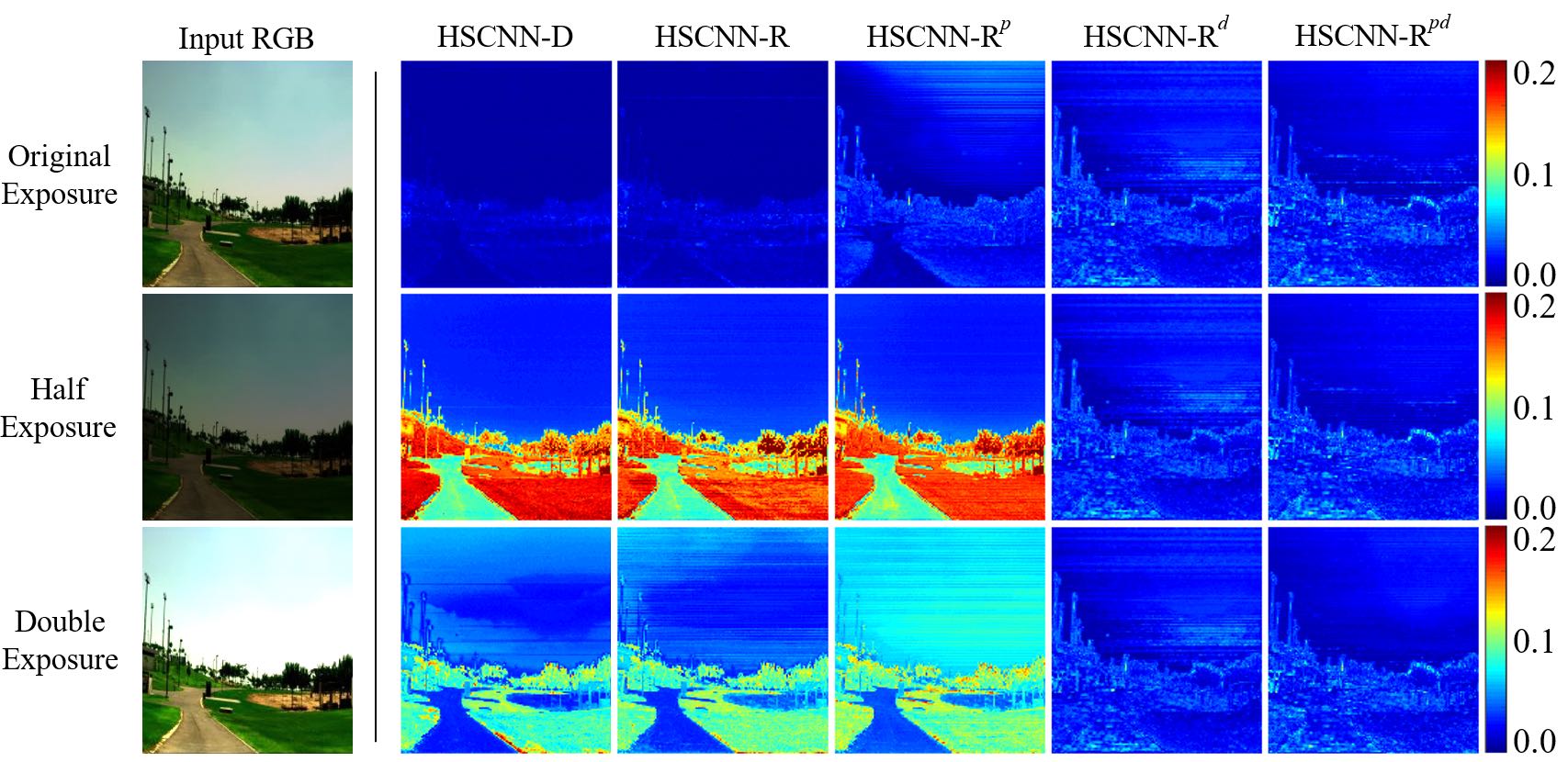}
    \caption{Visualization of spectral recovery errors by MRAE heat maps. All models are tested under original exposure (top row), half exposure (middle row) and double exposure (bottom row).}
    \label{fig:mrae_error_map}
\end{figure*}

First, we see that the state-of-the-art HSCNN-D and HSCNN-R are not physically plausible. Indeed, the worst-case $\Delta E$ of these models are significant (referring to \cite{sharma2002digital} human observers can sense noticeable difference above $\Delta E \approx 2.3$). Our physically plausible HSCNN-R$^p$ not only provides \textit{zero error} in color reproduction, but also significantly improves the worst-case performance in terms of spectral recovery over the original HSCNN-R. However, as shown in Figure \ref{fig:mrae_error_map}, HSCNN-R$^p$ and the original models provide poor spectral recovery performance when half and double exposure settings are applied.

\begin{figure*}[h!]
    \centering
    \includegraphics[width=1\linewidth]{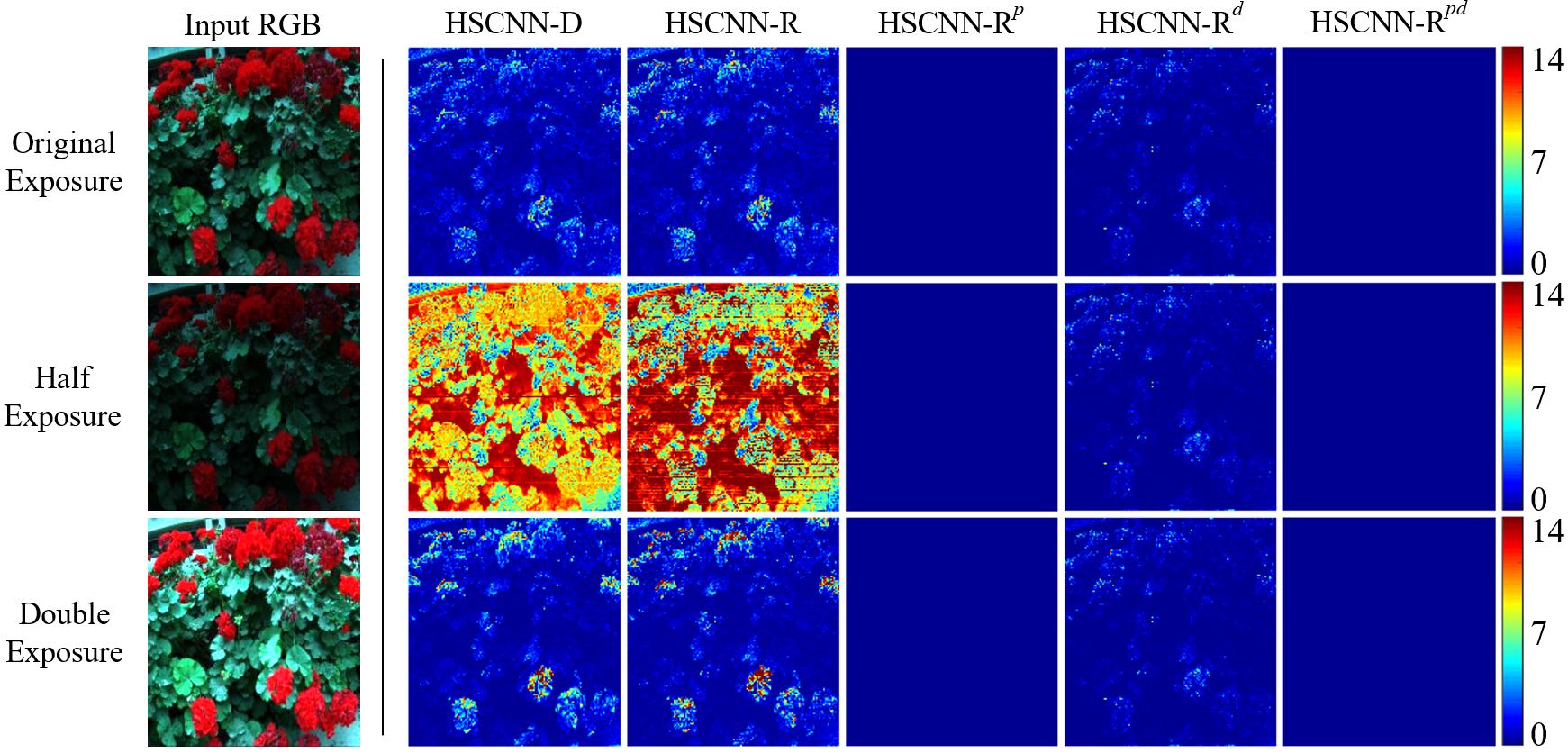}
    \caption{Visualization of color reproduction errors by $\Delta E$ heat maps. Referring to \cite{sharma2002digital} the threshold for human observers to notice the difference is around $\Delta E \approx 2.3$. All models are tested under original exposure (top row), half exposure (middle row) and double exposure (bottom row).}
    \label{fig:dE_error_map}
\end{figure*}

Next, we can see very clearly in Figure \ref{fig:mrae_error_map} and \ref{fig:dE_error_map} that standalone HSCNN-R$^d$ (without the physically plausible training) shows great advantage over original HSCNN-D and HSCNN-R in both the spectral recovery and color reproduction performance when exposure condition changes. However, the worst-case color reproduction performance of HSCNN-R$^d$ is still sub-optimal.

\begin{figure}[t]
\centering
    \begin{subfigure}[t]{0.9\linewidth}
        \centering
        \includegraphics[width=\linewidth]{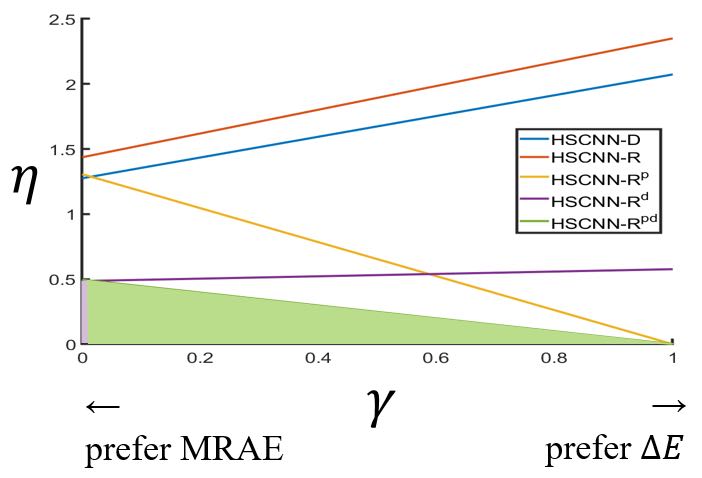}
    \end{subfigure}%
    ~\\
    \begin{subfigure}[t]{0.9\linewidth}
        \centering
        \includegraphics[width=\linewidth]{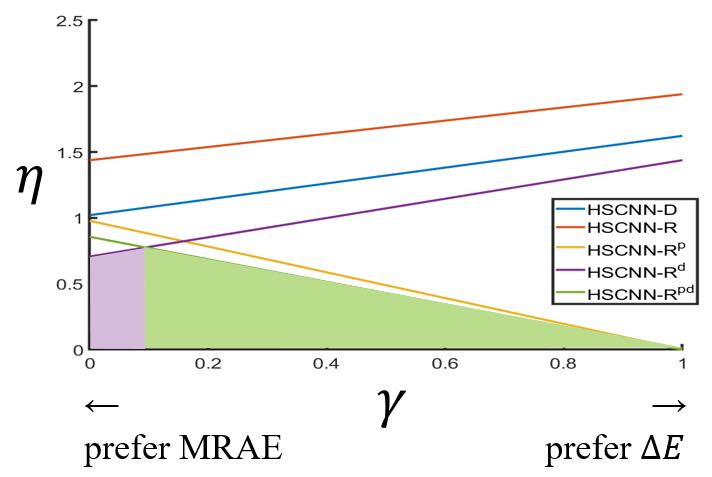}
    \end{subfigure}%
    \caption{Joint metric $\eta$ versus the relative weights $\gamma$ between MRAE and $\Delta E$. Respectively the top panel considers the mean and the bottom panel considers the worst-case MRAE and $\Delta E$ errors. The solid colored areas under the lowest curve indicate the best model at each $\gamma$.}
    \label{fig:joint}
\end{figure}

Lastly, we want to `jointly' consider the performance in spectral recovery and color reproduction. Frankly speaking, it is not possible to strictly say which performance is more important than the other. We also remark that depending on different applications, this relative importance can vary drastically. Despite of this, we can \textit{combine} the two metrics with an adjustable relative weight, to see the model performance in all different cases of relative importance. Define a joint metric $\eta$:
\begin{equation}
    \eta = \gamma \Delta E + (1-\gamma)\text{MRAE} \ ,
\end{equation}
the joint metric $\eta$ in effect describes the `competition' in importance between the two concerned metrics. We show two comparisons. First we show the mean MRAE against the mean $\Delta E$ in the top panel of Figure \ref{fig:joint}, and the worst-case MRAE competed with the worst-case $\Delta E$ in the bottom panel. Note that for each model we average the performances under the 3 testing exposure conditions, and we normalize all performances by the average performance across models (this is for making MRAE and $\Delta E$ in the same order of magnitude). We show that in both cases, either mean or worst-case performances, our proposed HSCNN-R$^{pd}$ performs the best overall.

\section{Conclusion}

Spectral reconstruction (SR) studies the mapping from RGB to hyperspectral images, which is regarded as a promising solution to low-cost, snapshot and high resolution hyperspectral camera. In the recent development of spectral reconstruction, leading models are based on Convolutional Neural Networks (CNN), providing remarkable spectral recovery performance. However, these models only aim to minimize the spectral recovery errors without ensuring the physical plausibility of the output spectra.
Physical plausibility is defined as ensuring the recovered spectrum integrates (using the underlying camera sensors) to the same RGB as it is recovered from. Existing method, which do not have this property, estimate RGBs which are significantly different from those found in the original image. 

In this paper we developed a physically plausible Spectral reconstruction framework.  Our insight is that all plausible spectra can be represented by a fixed camera-subspace projection spectrum defined by a linear combination of camera spectral sensitivities, and a \textit{null-space} spectrum which do not contribute to the color formation. Relative to this insight, the spectral recovery problem sets out to reconstruct the null-space spectra from the RGB (instead of the original RGB to radiance mapping), such that the physical plausibility of the predicted radiance is guaranteed. Finally, we also addressed the issue of exposure invariance in spectral reconstruction \cite{lin2019exposure}, by proposing a new data augmentation framework to ensure the model robustness against intensity variations. As the exposure changes, our models provide leading performance considering both spectral recovery and color reproduction.

{\small
\bibliography{egbib}
}

\end{document}

\title{Supplementary Material:\\
Spectral Reconstruction on Real RGB Data}

\maketitle

Most of the image-based spectral reconstruction (SR) methods are trained with \textit{simulated} RGB data (including, the experiments we present in the main paper), due to the difficulties of getting registered ground-truth hyperspectral and RGB images. But, presumably, we wish to use the SR models on real world data, where RGB images are taken by actual cameras. 

In this supplementary material we aim to show some visual evidence that the physical plausibility of SR is crucial for maintaining color fidelity in real-world applications.
More ambitiously, we wish to compare the actual `end-of-pipe' output of a camera system (\ie the processed image shown to the end users) with the prediction given by spectral reconstruction. In detail, with a model of the processing pipeline in hand (we shall introduce this model later), we (1) take a camera raw image, (2) recover the corresponding hyperspectral images by SR, (3) reintegrate the spectra with the given camera spectral sensitivities and finally (4) apply the pipeline model to this reintegrated raw image to generate an approximated end-of-pipe image.

\begin{figure}
    \centering
    \includegraphics[width=1\linewidth]{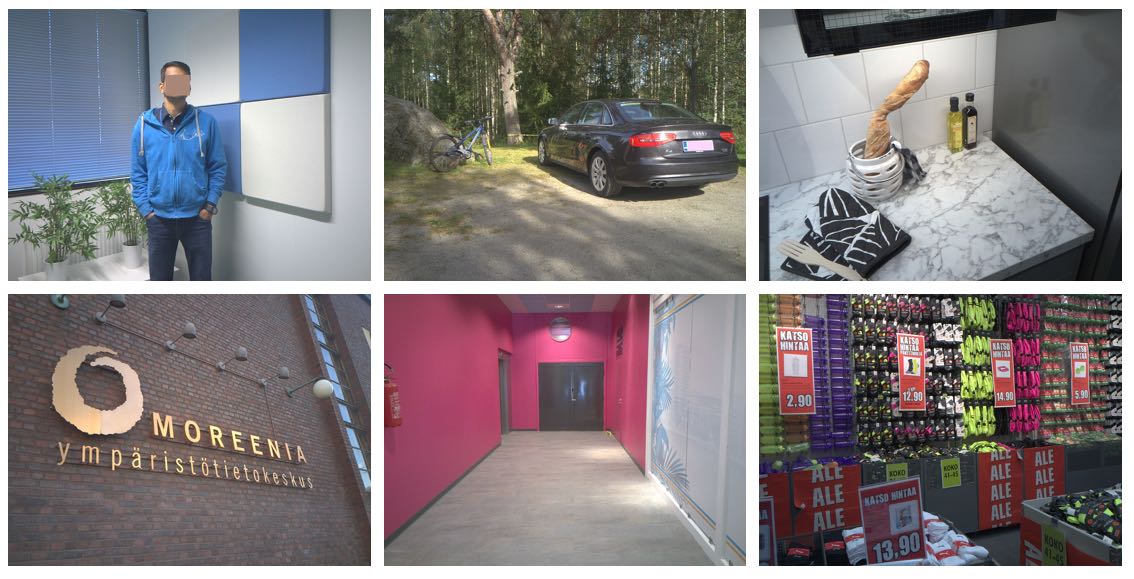}
    \caption{Example images in INTEL-TAU database \cite{laakom2019intel}}
    \label{fig:intel-tau}
\end{figure}

We get these `real RGB images' from the INTEL-TAU image database \cite{laakom2019intel}, which is by far the largest open-source database for training and evaluating the algorithms of color constancy (\ie illumination estimation). This database is very useful because it provides with each raw image: 
\begin{itemize}
    \item the spectral sensitivities of the camera used 
    \item the expected end-of-pipe rendered image
    \item the ground-truth white point color (WP)
    \item the color correction matrix (CCM) which maps the raw RGBs to sRGB colors.
\end{itemize}
We are going to make use of all the above information in our demonstration.
Example images from this database are given in Figure \ref{fig:intel-tau}.

\section{Training}

We re-trained two SR models for comparison: the original\footnote{The HSCNN-R model \cite{shi2018hscnn+} ranked the 2nd place in 2018 NTIRE Challenge on Spectral Reconstruction from RGB Images \cite{arad2018ntire}} HSCNN-R and the proposed HSCNN-R$^{pd}$ model (this second model is guaranteed to recover spectra that are colorimetrically accurate and also robust to variation in scene exposure). The purpose of this re-training is that we are to apply these models on real RGB data where the camera's spectral sensitivities are different from the CIE 1964 color matching functions \cite{commission1964cie} we used in the main paper.

Following the same training process as in the main paper, we randomly selected 100 ground-truth hyperspectral images from the ICVL dataset \cite{arad2016sparse} for training and 50 for validation (the spatial dimension of these images is around $1300\times 1392$). The only difference is that now the corresponding raw RGB images were simulated by the spectral sensitivities of SONY IMX135 (one of the three cameras used in INTEL-TAU). That is, the two SR models were trained to map SONY IMX135 raw RGBs to hyperspectral image output.

\section{Reconstruction}

The two trained SR models were used to reconstruct the hyperspectral information from 6 selected raw RGB images from the INTEL-TAU dataset \cite{laakom2019intel}, all of which were taken by SONY IMX135. The spatial dimension of these images is $2448\times 3264$. 

Then, the reconstructed hyperspectral images were again reintegrated into raw RGB images with the spectral sensitivities of SONY IMX135. At this stage, the proposed HSCNN-R$^{pd}$ is expected to give the exact same RGBs as the input raw RGB images, whereas HSCNN-R can generate different RGBs. The goal of this supplementary test is to visually demonstrate how different (how {\it wrong}) this recovery can be from a colorimetric point of view. 

\section{Color Fidelity Test}

From page 3 onward of this supplementary document, we are going to show several visual comparisons and quantitative error maps between the ground-truth and the SR predicted end-of-pipe RGB images. In this section we detail the image rendering process and the process of calculating the quantitative errors.

\subsection{Visual comparisons on end-of-pipe images}

\ \ \ 
In the processing pipeline of a camera, the raw image might undergo, but not limited to, the following processes before being shown to the end users: black level and saturation correction, white balancing, color correction and gamma correction. 
As we are already given the expected
end-of-pipe image with each raw image in the INTEL-TAU database, we can alternatively build a 3D Look-up-table (LUT) which approximates the actual image processing pipeline: for each image, the LUT is built 
to relate each color in the ground-truth raw RGB image to the colors in the supplied (expected) end-of-pipe image.

This LUT can be optimized - in a least-squares sense - by \textit{lattice regression} \cite{lin2012nonuniform,garcia2009lattice}. 
To speed up the optimization process, we train the LUT on \textit{thumbnail images}, where we simply downsample the images from the original $2448\times 3264$ to $108\times 144$, and bin the colors by $24\times 24\times 24$ in the three color channels. Then, the full resolution ground-truth raw RGB and the raw RGB reintegrated from the reconstructed hyperspectral image are mapped to their respective end-of-pipe renditions by applying the same 3D LUT.

In Figure 3-8, an example image is shown in the bottom-left of each figure, in which the 4 regions of interest are marked with white squares. The `Ground Truth' image (top-left of each figure) is actually the end-of-pipe image rendered by the trained 3D LUT mapping. On the other hand, from the ground-truth raw RGB we carry out spectral reconstruction (\ie the two trained SR models) and reintegrate the recovered hyperspectral images with the camera sensitivities to get an {\it approximate} raw image. By applying the same LUT to this derived raw image we generate the end-of-pipe images predicted by the two SR models, as shown in the top-middle and top-right images in each figure. 

We can already see that HSCNN-R, as an physically non-plausible spectral reconstruction model, introduces color shifts that are quite visible after color rendering, while our physically plausible HSCNN-R$^{pd}$  successfully preserves the original colors in the ground-truth images. To further quantify the color shifts, we are bound to calculate the \textit{color difference} between the ground-truth and the reintegrated RGB images.

\subsection{Quantifying color differences}

\ \ \ We wish to use the CIE 1976 color difference ($\Delta E$) \cite{robertson1977cie} to quantify the colorimetric errors. Since the $\Delta E$ is defined in CIELAB color coordinates (as shown in Equation (19) in the main paper), we must consider how we transform the camera raw RGB to their CIELAB counterparts.

\begin{figure}
    \centering
    \includegraphics[width=1\linewidth]{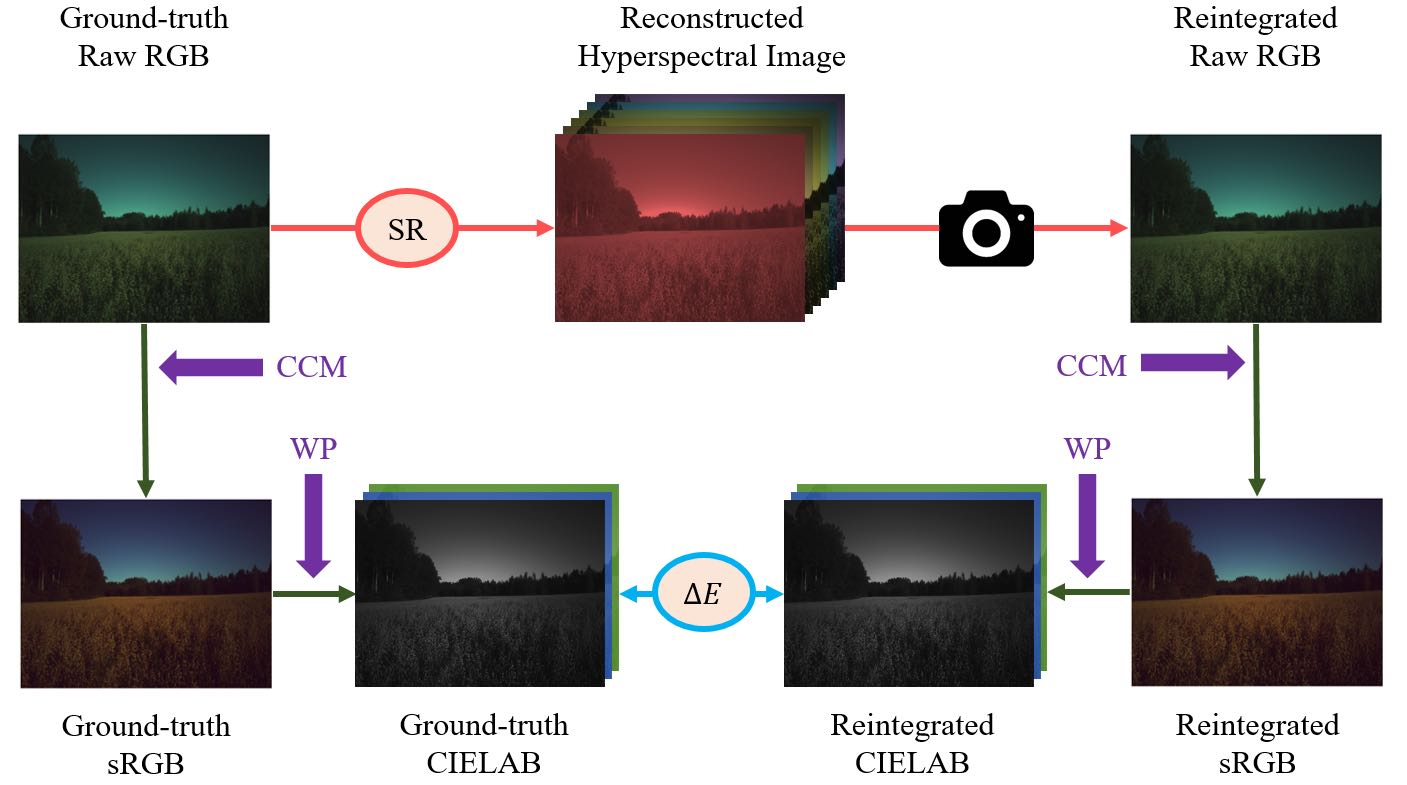}
    \caption{The process of calculating CIE 1976 color difference $\Delta E$ between ground-truth and reintegrated color images.}
    \label{fig:color_transformation_flowchart}
\end{figure}

The procedure is summarized in Figure \ref{fig:color_transformation_flowchart}. Unlike in the main paper where the CIELAB coordinates can be transformed directly from the CIEXYZ colors (with the white point color this mapping is one-to-one \cite{susstrunk1999standard}), the mapping from the real camera's raw RGB to CIELAB is \textit{unknown} if the raw data is the only given information.
Fortunately, INTEL-TAU also provides with each raw image the \textit{color correction matrix} (CCM) that transforms the image into sRGB colors and the information of ground-truth white point (WP) that ensures one-to-one mapping between sRGB and CIELAB \cite{susstrunk1999standard}. Finally, the desired $\Delta E$ color difference between ground-truth and reintegrated RGB images can be calculated from the transformed CIELAB images.

We show the $\Delta E$ error maps in the bottom-middle and bottom-right of Figure 3-8, which detail the pixel-wise colorimetric errors introduced by the two trained SR models in the 4 selected regions of interest. It is evident that HSCNN-R recovers spectra that reintegrate into \textit{wrong} colors with significant errors (we remark once again that referring to \cite{sharma2002digital} human observers can sense noticeable color difference above $\Delta E \approx 2.3$). Remarkably, our proposed HSCNN-R$^{pd}$ model - which possesses both physical plausibility and exposure invariance - preserves complete color fidelity.

\begin{figure*}
    \centering
    \includegraphics[width=0.9\linewidth]{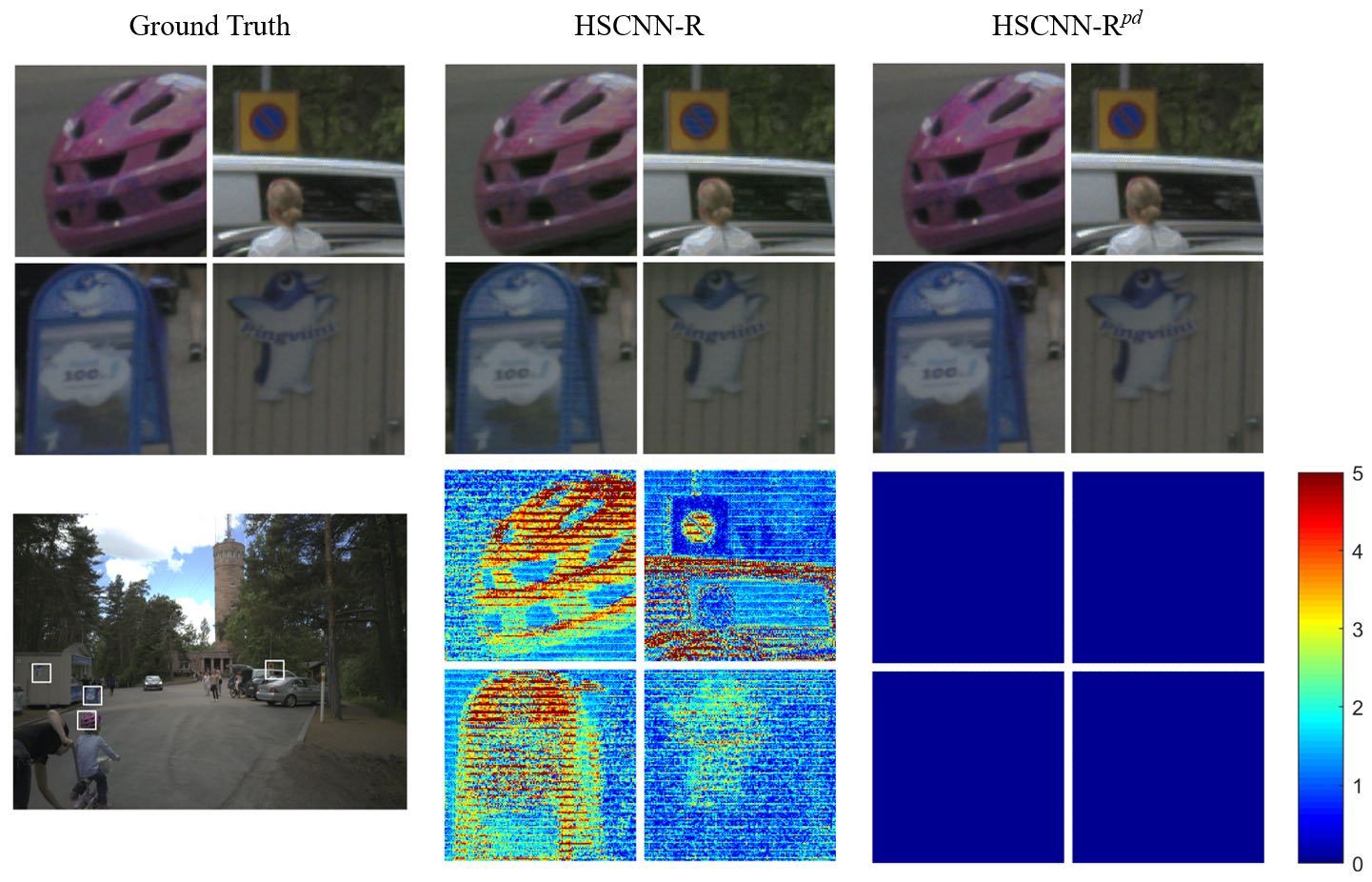}
    \caption{Visual comparison 1. Top row: the rendered images. Bottom row: the corresponding $\Delta E$ error maps.}
\end{figure*}
    
\begin{figure*}
    \centering
    \includegraphics[width=0.9\linewidth]{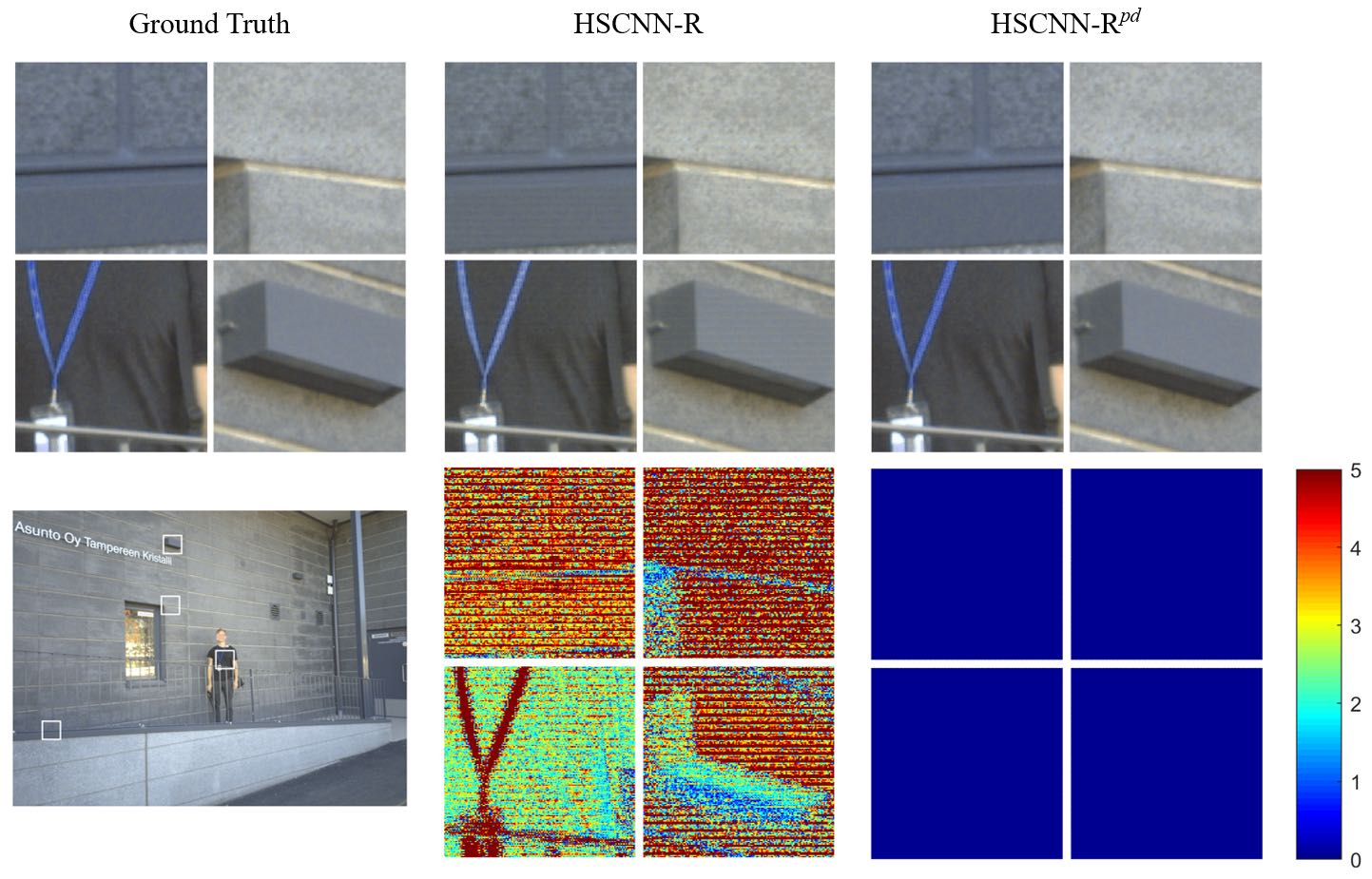}
    \caption{Visual comparison 2. Top row: the rendered images. Bottom row: the corresponding $\Delta E$ error maps.}
\end{figure*}

\begin{figure*}
    \centering
    \includegraphics[width=0.9\linewidth]{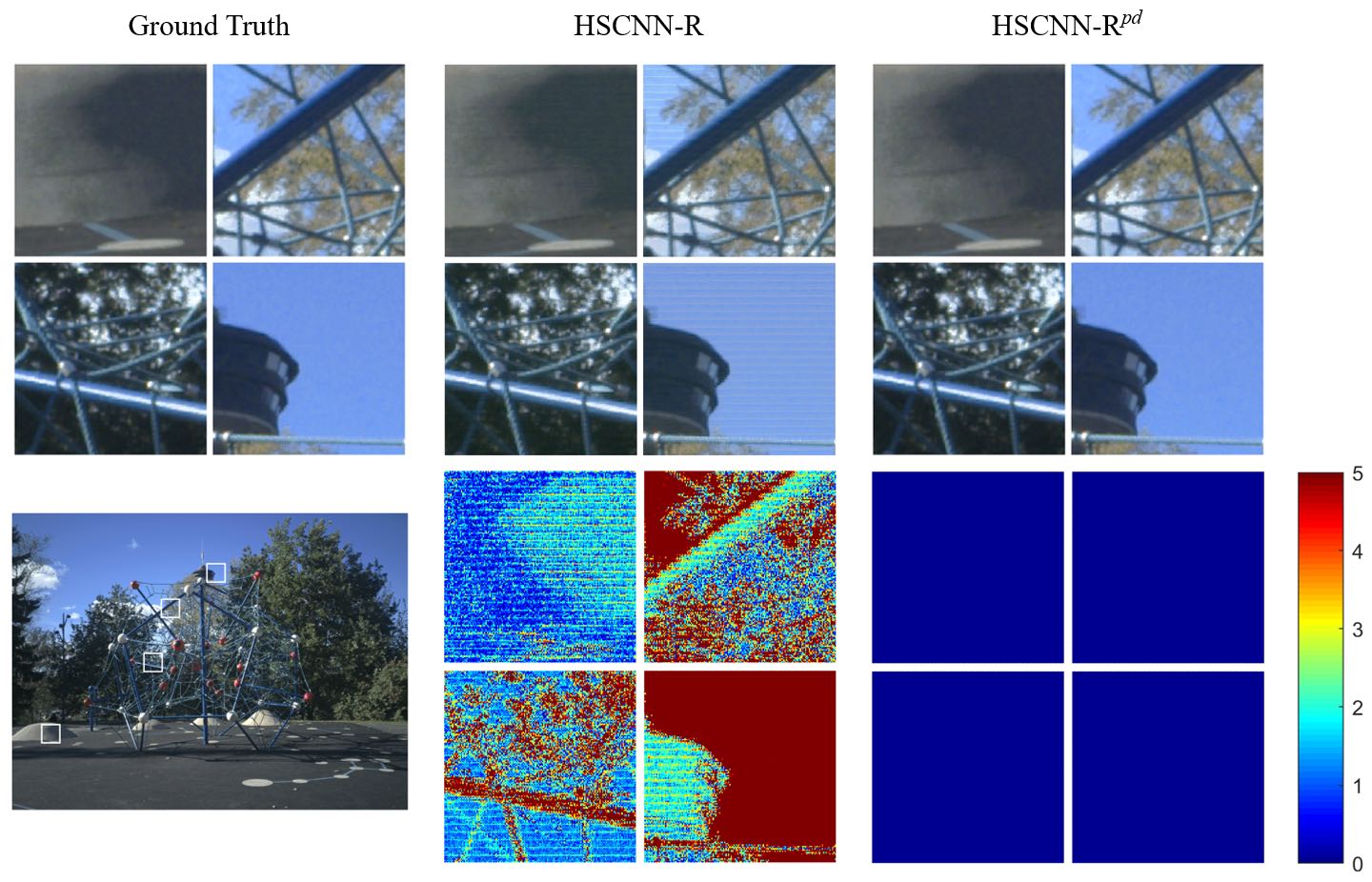}
    \caption{Visual comparison 3. Top row: the rendered images. Bottom row: the corresponding $\Delta E$ error maps.}
\end{figure*}
    
\begin{figure*}
    \centering
    \includegraphics[width=0.9\linewidth]{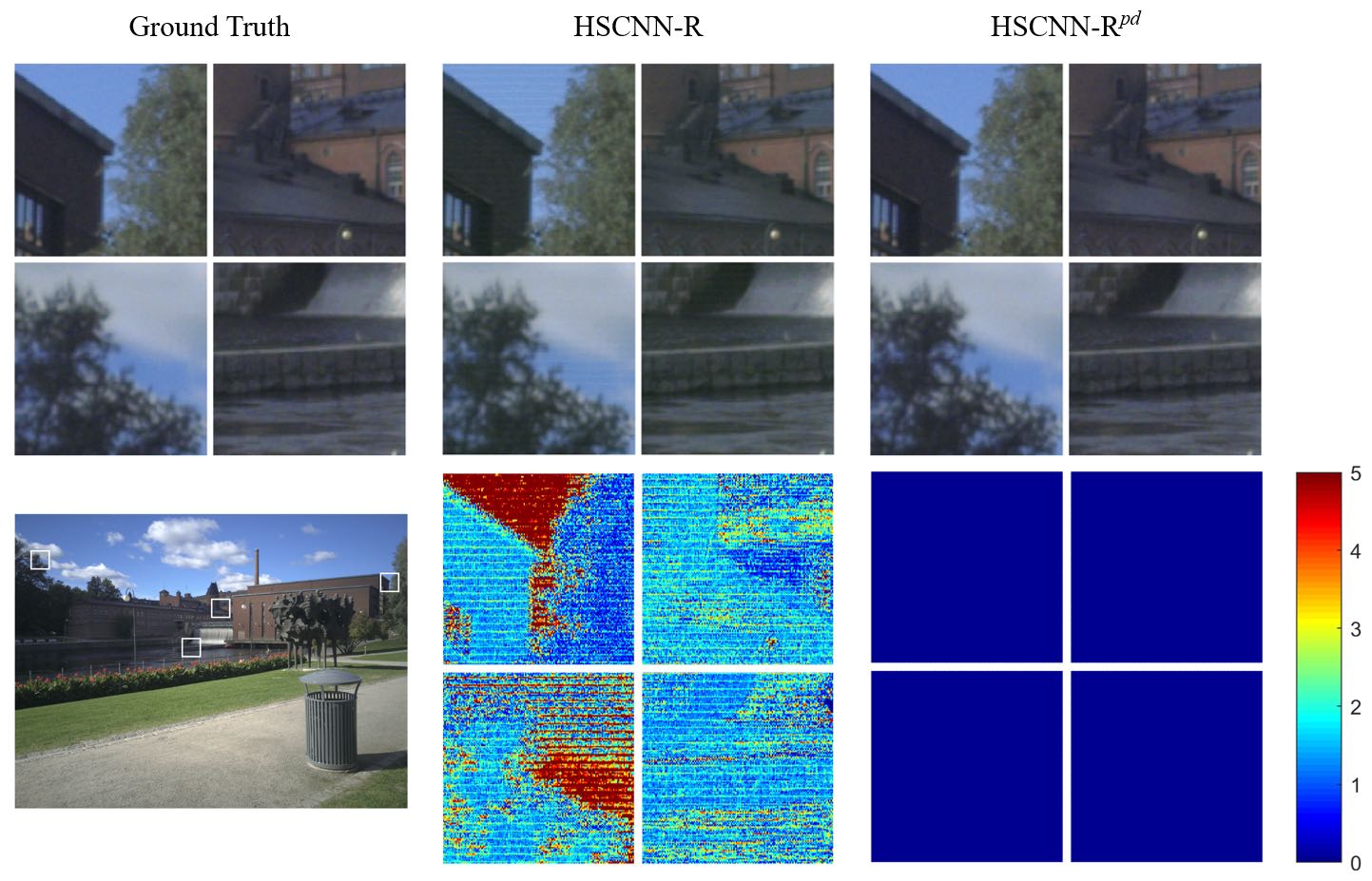}
    \caption{Visual comparison 4. Top row: the rendered images. Bottom row: the corresponding $\Delta E$ error maps.}
\end{figure*}

\begin{figure*}
    \centering
    \includegraphics[width=0.9\linewidth]{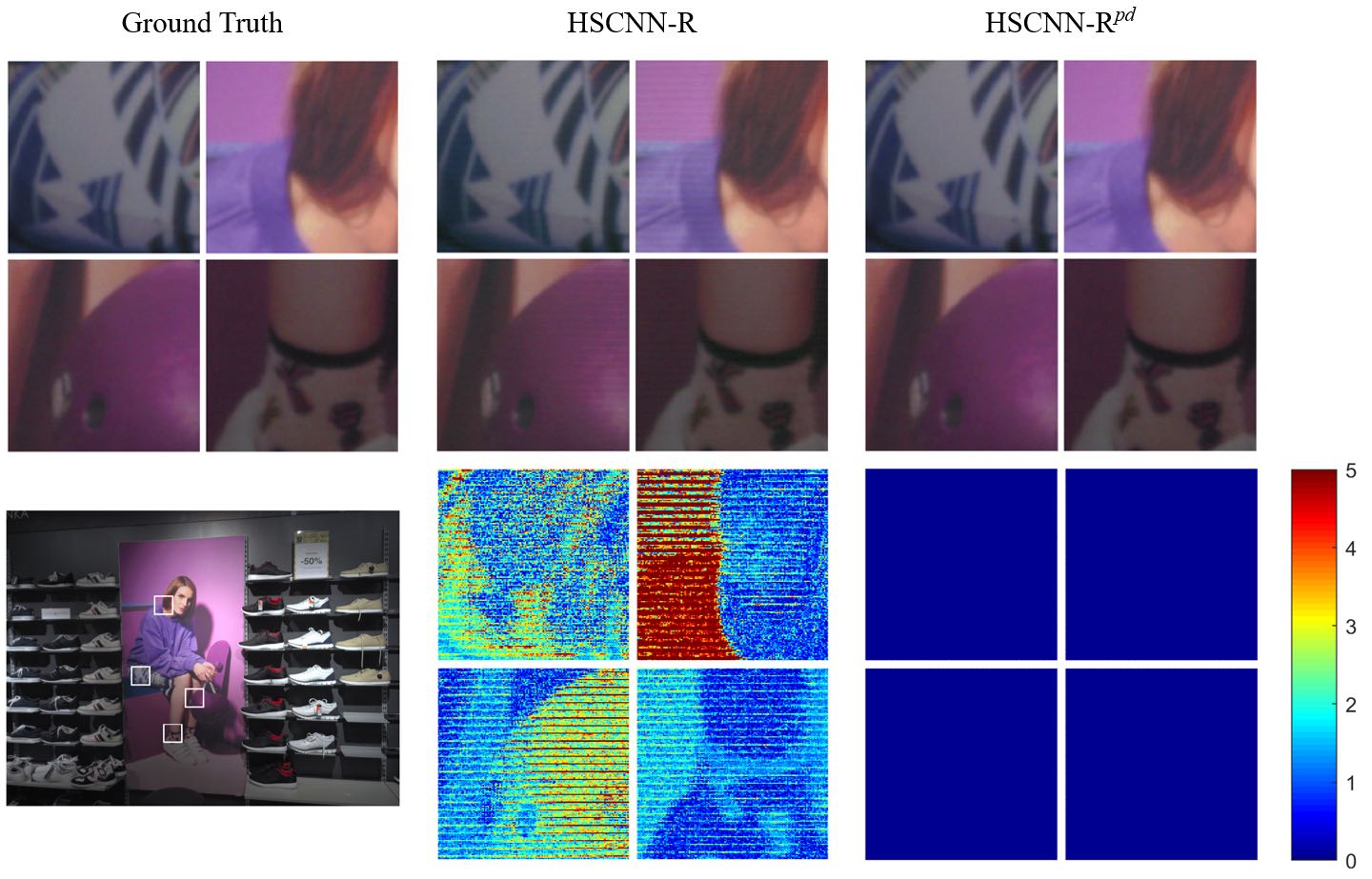}
    \caption{Visual comparison 5. Top row: the rendered images. Bottom row: the corresponding $\Delta E$ error maps.}
\end{figure*}
    
\begin{figure*}
    \centering
    \includegraphics[width=0.9\linewidth]{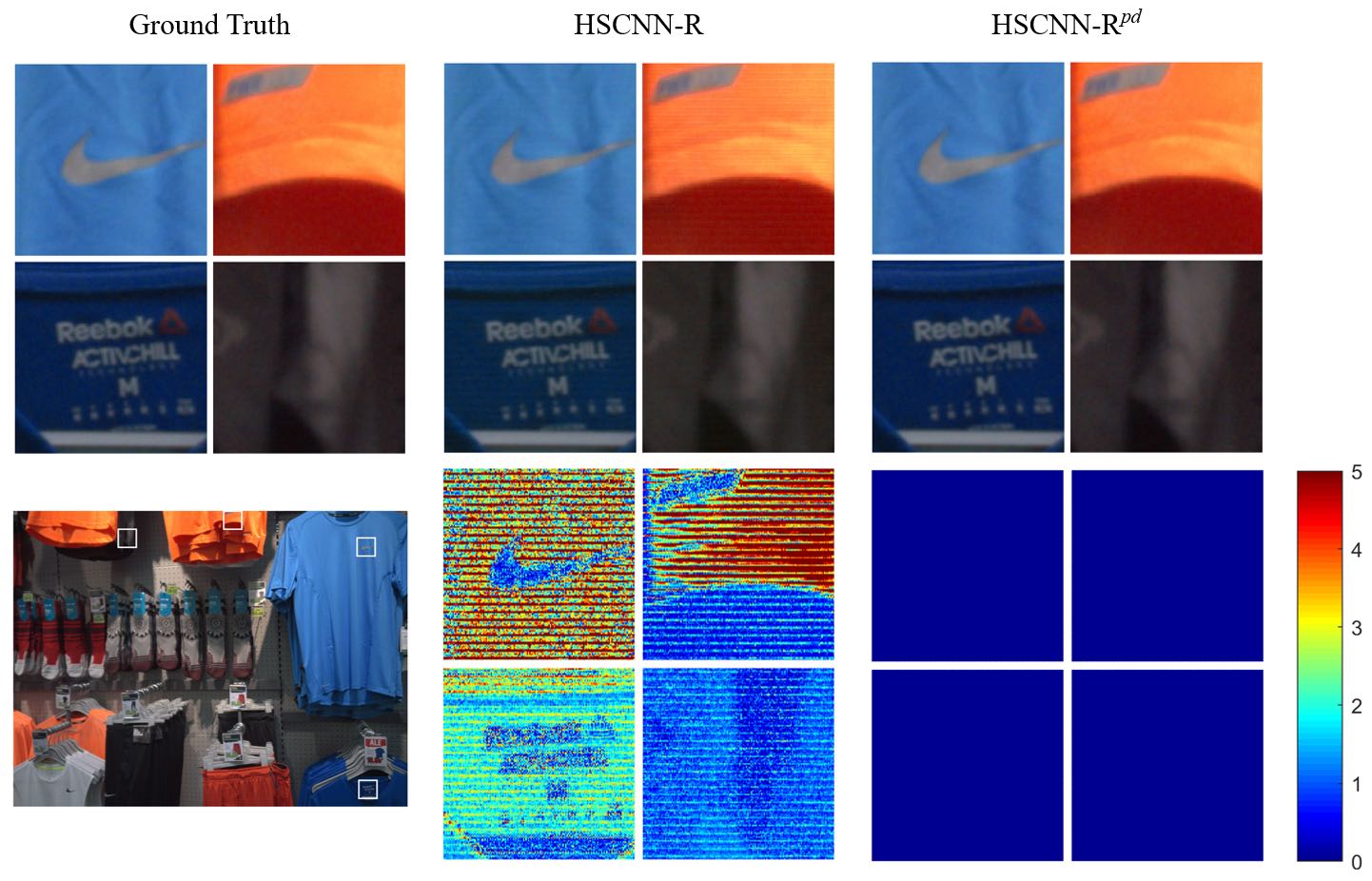}
    \caption{Visual comparison 6. Top row: the rendered images. Bottom row: the corresponding $\Delta E$ error maps.}
\end{figure*}

{\small
\bibliography{egbib}
}